\documentstyle[prl,aps,amsmath,multicol,epsfig]{revtex}

\newcommand{\bsigma}{\boldsymbol{\sigma}}
\newcommand{\soq}{S(|\mathbf{q}|)}
\draft
\begin{document}

\title{Phonons from neutron powder diffraction}

\author{D. A. Dimitrov, D. Louca and H. R\"oder}
\address{Los Alamos National Laboratory, Los Alamos, NM87544}
\date{\today}

\maketitle
\begin{abstract}
The spherically averaged structure function $\soq$ obtained from pulsed neutron
powder diffraction contains both elastic and inelastic scattering via an integral
over energy. The Fourier transformation of $\soq$ to real space, as is done in the
pair density function (PDF) analysis, regularizes the data, i.e. it accentuates the
diffuse scattering.
We present a technique which enables the extraction of  off--center
($|\mathbf{q}| \neq 0 $) phonon information from powder diffraction 
experiments by comparing the experimental
PDF with theoretical calculations
based on standard interatomic potentials and the crystal symmetry. 
This procedure (dynamics from powder diffraction(DPD)) 
has been \emph{successfully}  implemented for two systems,
a simple metal, fcc Ni, and an  ionic crystal, CaF$_{2}$.
Although computationally intensive, this data analysis allows for a
phonon based modeling of the PDF, and additionally
provides  off--center phonon information from powder neutron diffraction. 
\end{abstract}
\begin{multicols}{2}
\narrowtext
For a variety of physical questions it is significant to obtain information on
off-center phonons in crystals. This is particularly important in complex materials
where local effects modify the macroscopic properties, as for example in
 high temperature superconductors, colossal magneto--resistive materials, ferroelectrics,
intermetallic alloys and many more. Until this study the only available method to
obtain off--center phonon data  has been 
inelastic neutron scattering, which is intensity
limited, hence time consuming, and for detailed studies, 
relies on the availability of large single
crystals (triple axis measurements). 
Here we show how to
obtain similar data from powder neutron diffraction.

Historically,
the purpose of powder (polycrystalline) neutron diffraction has been the
exact determination of the average crystal structure using modern crystallographic
analysis techniques like Rietveld refinement, see e. g. \cite{rietveldbook}. Within this
type of refinement only a limited range of the momentum transfer $q$ is necessary.
More recently, however,
the availability of pulsed sources has made  possible the measurement of 
$\soq$ up to  very large values of $q$. 
The PDF analysis \cite{Egami90} 
has made use of this additional information
to investigate local atomic deviations
from an average crystallographic structure.

To model the peak positions in the PDF analysis one calculates  
interatomic distances from the crystallographic
unit cell. The peak shape
is commonly fitted to the experimentally obtained PDF, $\rho^{exp}(r)$, 
 using Gaussians.
Since the experimental PDF contains additional information from the diffuse
scattering, observed differences have  been successfully attributed 
to local structural deformations
like e. g. polarons \cite{Louca97}. 
This type of modeling of the PDF does not take into account the
intrinsic peak widths
caused by coherent excitations in solids like phonons or spin waves, and
how the peak widths are modified by a  finite momentum transfer cut--off.
Other attempts to relate the measured PDF peak widths 
to the intrinsic phonon dynamics
in a real space approach \cite{Chung97}
were not intended to address the inverse problem,
i. e. extracting the phonon parameters from the PDF.
In a real space approach, it is considerably more difficult 
to take into account possible extinction
rules for inelastic scattering caused by the 
point group symmetry of the reciprocal
lattice, as has been recently shown \cite{dododo}. Also in reciprocal space
the instrumental resolution function is  easier taken care of.
In addition to information on lattice dynamics we provide
an extended modeling of the PDF based on the estimated phonon parameters.

Triple axis neutron scattering is the most direct way to measure phonons.
Indirect measurements on powders are either based on 
a detailed analysis of the peak shape in $\soq$, as is done 
in the analysis of thermal diffuse scattering (TDS) \cite{Schofield87},
or via a reconstruction of $S(|{\bf q}|,\omega)$ from  time--of--flight (TOF)
experiments, see e.g. \cite{Schoberstuff}. Inelastic
experiments are intensity limited, and the TDS analysis relies on a very
good representation of the background. 
In this letter, we
show how to extract information on off--center phonons  in a
reliable way from $\soq$ determined from TOF experiments. 
This is  achieved via a \textbf{parameterization} 
of the phonon dispersion curves
and eigenvectors using a suitable model (as is standard in the presentation of
triple axis data, see e.g. \cite{Biltzbook,Landolt}), from which we can 
calculate a theoretical $\soq$. The data are
then \textbf{regularized} by transforming to real space 
to obtain a theoretical PDF, $\rho^{theo}(r)$. A reverse Monte
Carlo procedure is then 
used to \textbf{estimate} the parameters from a comparison with
the experimental data, $\rho^{exp}(r)$,  and to give data driven error bars.

The measured quantity  is  
 the powder (angular) averaged structure factor 
$ S(|\mathbf{q}|) $
(normalized by the neutron scattering lengths). For simplicity we assume that
the incoherent effects can be neglected, and decompose the total structure factor 
into elastic and one--phonon
inelastic contributions: 
$ S^{tot}_{coh}({\bf q})\approx
 S_{coh}^{elastic}({\bf q})+S_{coh}^{1}({\bf q}) $.
Note that $ S^{tot}_{coh}(\mathbf{q}) $ contains all dynamic information
as an integral over frequency\cite{footnote1}.
 The elastic part is given by \cite{Loveseybook}
\begin{equation}
\label{elasticxsection}
S_{coh}^{elastic}({\bf q})=\frac{(2\pi )^{3}}{N_{b}v_{0}}\frac{1}{\overline{b}^{2}}\sum _{{\bf G},d}\delta ({\bf q}-{\bf G})\overline{b}_{d}e^{i{\bf q}.{\bf d}-W_{d}({\bf q})}
\end{equation}
 and the one--phonon inelastic part by
\begin{eqnarray}
S_{coh}^{1}({\bf q}) & = & \frac{\hbar (2\pi )^{3}}{2N_{{\bf k}}N_{b}v_{0}}\frac{1}{\overline{b}^{2}}\sum _{{\bf G}}\sum _{{\bf k}\in BZ,j}\left| 
F_j({\bf q})
\right| ^{2}
\notag \\
 & \times  & \delta ({\bf q}-{\bf G}+{\bf k})\frac{\coth (\frac{\beta }{2}\hbar \omega _{_{j}}({\bf k}))}{\omega _{j}({\bf k})}\; ,\label{onephonon} 
\end{eqnarray}
where $F_j({\bf q})=\sum _{d}\frac{\overline{b_d}}{\sqrt{M_{d}}}({\bf q}.\bsigma ^{j}_{d}({\bf k}))e^{-W_{d}({\bf q})+i{\bf q}.{\bf d}}$,
the sum over $ \mathbf{G} $ runs over all vectors of the reciprocal
lattice, $ \mathbf{k} $ describes a Brillouin zone integral over $ N_{{\bf k}} $
points, $ d $ runs over atoms in the cell and $ j $ counts the branches.
$ \overline{b_{d}} $ is the average scattering length and $ M_{d} $ the
mass of the $ d $'th atom, $ v_{0} $ the unit cell volume, and $ W(\mathbf{q}) $
is the Debye-Waller factor defined as 
%
$W_{d}({\bf q})=\frac{\hbar }{4N_{k}M_{d}}\sum _{{\bf k}\in BZ,j}\frac{\left| {\bf q}.\bsigma ^{j}_{d}({\bf k})\right| ^{2}}{\omega _{j}({\bf k})}\coth (\frac{\beta \hbar }{2}\omega _{j}({\bf k}))\; 
$.
The phonons enter eqns (\ref{elasticxsection}) and (\ref{onephonon}) via the dispersion
relations $ \omega _{j}(\mathbf{k}) $ and the eigenvectors $ \bsigma _{d}^{j}(\mathbf{k}) $.
The powder averaged structure function $ S(|{\bf q}|)\equiv \frac{1}{4\pi }\int _{|{\bf q}|=const}d\Omega _{{\bf q}}\: S(\mathbf{q}) $
must be obtained numerically, and is CPU time consuming for the inelastic
part of eqn(\ref{onephonon}). To compare to the experimental situation it is 
necessary to convolute $\soq$ 
with a suitably chosen resolution function\cite{Egami90} which depends on the instrument.
\begin{figure}[h]
\label{nirho}
\caption{The experimental $\rho^{exp}(r)$ compared with the theoretical $\rho^{theo}(r)$ 
calculated for the expectation values 
\protect{$\{ \overline{f_i}\}$} resulting from the DPD
procedure for Ni at room temperature. $\sigma(r)$ are the experimental errors in the
$\chi^2$ estimator (\protect\ref{chisquared}).}
\epsfig{file=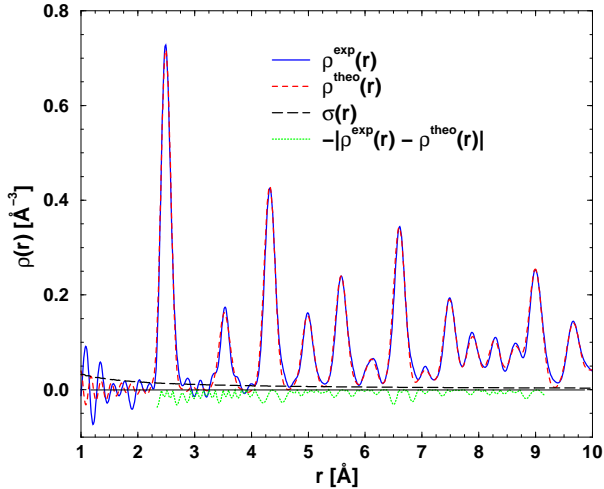,height=3.1in,angle=270}
\end{figure}
%
The extraction of the lattice dynamics described by
$ \omega _{j}({\bf q})$ and $\bsigma _{d}^{j}({\bf q}) $ using  a comparison
of TOF  experimental data  with a theoretical 
$\soq$ is difficult in reciprocal space
because $\soq$ is a superposition of elastic
scattering, i.e. the Bragg peaks, 
 and inelastic scattering,
 both of which are vastly different in amplitude.
A direct comparison in reciprocal space 
is further complicated by the unavoidable instrument background
which is superimposed on the data. 
Both problems can be greatly reduced by transforming
$ \soq  $ into the real space pair density function $ \rho (r) $ via Fourier
transform
\begin{equation}
\label{rhoofq}
\rho (r)=\rho _{0}+\frac{1}{2\pi ^{2}r}\int _{0}^{\infty }dq\: \left[ S(|{\bf q}|) -1\right] q\sin (qr)\; .
\end{equation}
Pulsed sources provide a high enough momentum transfer to reduce the truncation 
error in eqn (\ref{rhoofq}) and in turn, allow for a proper normalization of the data.
Due to the properties of a Fourier transform the peaks in $ \rho (r) $ are
all of comparable height 
which eliminates the problem
with the large amplitude fluctuations in $ \soq  $, i.e. it regularizes the
data analysis problem. As a useful byproduct, all features in $ \soq  $ with inverse
length scales larger than $ \frac{2\pi }{a_{0}} $, i.e. most of the background,
are transformed into distances in $ \rho (r) $, which are smaller than the
shortest interatomic distance $ a_{0} $, and can therefore be neglected in the parameter
estimation. 
\begin{figure}[h]
\label{nidisp}
\caption{
The phonon dispersion curves for room temperature Ni. The symbols are 
triple axis data from \protect\onlinecite{Birgeneau64}. 
The shaded bands  are the
one--sigma error intervals of the dispersion curves obtained by our procedure. 
The green bands result from just reproducing the PDF, and the red bands
are from constraining the reverse Monte Carlo to agree with the measured
elastic constants within their error bars.}
\epsfig{file=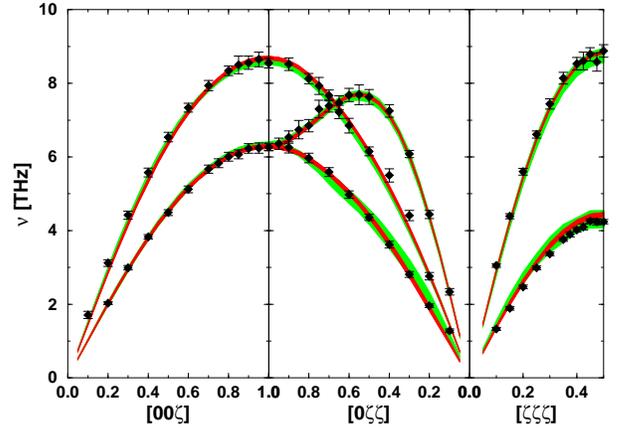,height=3.1in,angle=270}
\end{figure}
We use a $\chi^2$--functional in the parameter estimation process defined by
\begin{equation}
\chi ^{2}\left( \{f_{i}\}\right) =\int _{r_{a}}^{r_{b}}dr\: 
\frac{\left[ \rho ^{exp}(r)-\rho ^{theo}(r;\{f_{i}\})\right] ^{2}}{\sigma (r)^{2}}\; 
\label{chisquared}
\end{equation}
where $ \sigma (r) $ is the error in $ \rho ^{exp}(r) $, which
can be assumed to be $ \frac{\sigma _{0}}{r} $ \cite{Egami90}. The variational
parameters are 
the parameters describing
the lattice dynamics $ \{f_{i}\}, $ which depend on a theoretical model (see
below for concrete examples). $ r_{a} $, $ r_{b} $ define the range over
which the PDFs are compared.
We use a Monte Carlo procedure \cite{McGreevy95}
which gives statistical estimators for the parameters $ \overline{f_{i}} $,
and also error bars $ (\Delta f_{i})^2\equiv 
{\overline{f_{i}^{2}}-\overline{f_{i}}^{2}} $.
The $ \Delta f_{i} $ are controlled by the quality of the phonon model (a
bad model also giving a large $ \chi ^{2} $, i.e. systematic error), 
and by the quality of the data, the statistical error.
Due to the nature of our data, the experimental PDF, the dynamics is obtained
by averaging over the whole reciprocal space, not just along symmetry
directions, as is usually done, when triple axis data are fitted.
%
%
\begin{figure}[h]
\label{caf2rho}
\caption{The same as Fig. 1 but for $CaF_2$ at room temperature.}
\epsfig{file=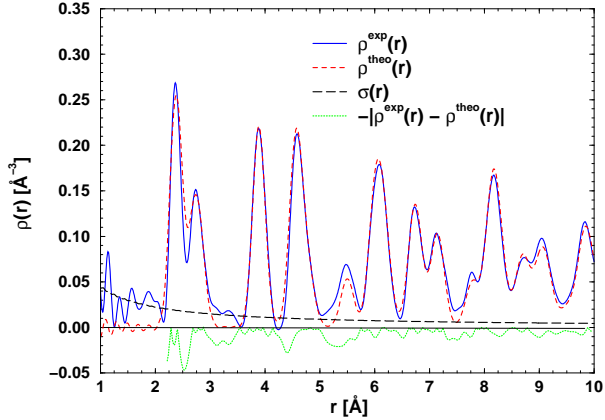,height=3.1in,angle=270} 
\end{figure}
%
The parameterization of the lattice dynamics involves the choice of a theoretical
model. There exists a large variety of models which describe
the measured dispersions from triple axis data \cite{Biltzbook,Landolt}.
To validate our procedure we chose to perform the above analysis on two
materials with very different models for the lattice dynamics. The first example
is fcc Ni, whose phonon dispersion 
can be described by a simple force constant model
\cite{Birgeneau64,DeWit68}. In Fig. 1,
the PDF determined from the diffraction data is compared with the 
PDF calculated from the estimated parameters.
The agreement is excellent and indicates
that this procedure does indeed reproduce \textbf{both} the peak position and peak
shape of a PDF measurement. 
A comparison of the force constants resulting from our analysis with  existing
triple axis data is shown in Table 1.
Given the force constants and their standard deviations  we can easily calculate
the resulting phonon dispersion curves. In Fig. 2  we show those
curves together with published triple axis data \cite{Birgeneau64}. 
Again the agreement is excellent proving that
we managed to extract the dynamical information contained in $ \soq  $.
Adding more information, e.g elastic constants, to the functional 
(\ref{chisquared}) reduces the errors in our parameter estimation. 
The additional constraints
lead to smaller error bars, and a slight improvement in the overall shape
of the dispersions.

As a second example we took $ CaF_{2} $, since it involves a fairly ionic
system, which necessitates a different interatomic force model. 
Also the phonons are a lot more 
complicated due to the presence of optical branches.
The phonons were parameterized by ionic shell models, as described in
Elcombe \& Pryor \cite{Elcombe70}. This is a more challenging problem as we now have
a much higher dynamic range (phonon energies up to $ 14$THz ), and the optical
modes show an unusually large dispersion. The results of the DPD
 analysis are presented
together with some available triple axis data in Table 2 for the force constants,
and in Figs. 3  and 4 for the comparison of PDFs and dispersions, respectively.
Again, the agreement is remarkable and shows that we can obtain reliable off-center
phonon information even for  non-trivial systems.
\begin{figure}[h]
\label{caf2disp}
\caption{
The phonon dispersion curves for $CaF_2$ at room temperature. The symbols are 
measured data from [16].
The shaded bands  are
one--sigma intervals of the dispersion curves obtained by our procedure. 
.}
\epsfig{file=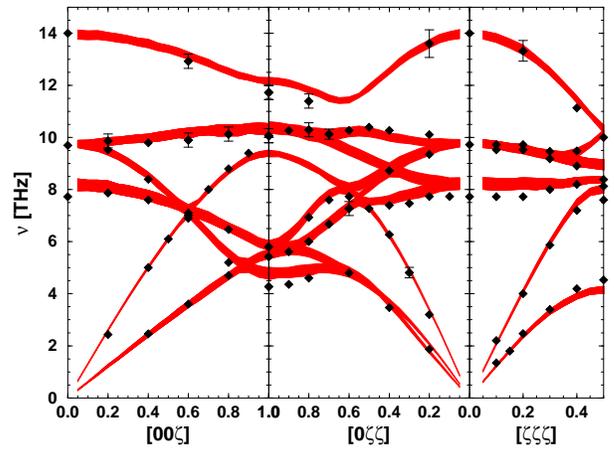,height=3.1in,angle=270}
\end{figure}
In this letter we have shown how dynamical information  present in
$\soq$ data can be extracted. The success of the present
 analysis relies on high quality data and on intelligent modeling
of the lattice dynamics. Noisy data or bad normalization will result in large
error bars. The models can easily be extended and modified
if the error analysis indicates through an unusually large $ \chi ^{2} $-value
that the model is incompatible to the data. Such extensions may include
more shells, or a totally different modeling based on pseudopotential calculations.
As additional constraints to the functional in eqn.(\ref{chisquared}) we can use further
data from zone center experiments, like elastic constants and Raman data. At
very low temperatures where the one-phonon processes are not thermally activated, the
information is not in the data. If necessary, more sophisticated Placzek corrections
and the incoherent scattering
can be included in the analysis. Our results are very encouraging, and we will
pursue this effort on more complicated materials, where triple axis data are
sparse or non-existent.
\acknowledgements
We would like to thank D. Wallace, J. Wills, A. R. Bishop, 
T. Egami, R. Silver, G. Straub, A. Lawson, and
G. H. Kwei for helpful discussions and support.
The Intense Pulsed Neutron Source (IPNS) of Argonne National Laboratory is
supported by the U.S. Department of Energy, Division of Materials Sciences,
under contract W-31-109-Eng-38. Work at the Los Alamos National Laboratory is performed under the auspices of the U.S. Department of Energy under contract W-7405-Eng-36.

%
%
%
%
\begin{table}
\caption{Generalized force constants $f\pm\Delta f$ for Ni 
	at room temperature (in $10^4$ dyn/cm) from reverse 
	Monte Carlo runs with (a) the elastic constants constraints,
	(b) without , and from
	a triple-axis measurement Ref.~\protect\onlinecite{DeWit68}.
	The errors in the force constants in 
	Ref.~\protect\onlinecite{DeWit68} are estimated at about 
	$\pm 0.04\times 10^4$ dyn/cm.
	The notation follows
	Ref.~\protect\onlinecite{DeWit68}.	
	}
\label{table_1}
\begin{tabular}{cr@{}l@{${}\pm{}$}r@{}lr@{}l@{${}\pm{}$}r@{}lr@{}lr@{}l}
  & \multicolumn{4}{c}{(a)} & 
      \multicolumn{4}{c}{(b)} &
        \multicolumn{2}{c}{Ref.~\protect\onlinecite{DeWit68}}\\
\tableline
 1XX  & 1&.755&  0&.018 & 1&.683 & 0&.024 & 1&.7319\\
 1ZZ  &-0&.054&  0&.019 &-0&.039 & 0&.019 &-0&.0436\\
 1XY  & 1&.878&  0&.017 & 1&.869 & 0&.159 & 1&.9100\\
 2XX  & 0&.067&  0&.023 & 0&.122 & 0&.046 & 0&.1044\\
 2YY  &-0&.011&  0&.008 &-0&.085 & 0&.031 &-0&.0780\\
 3XX  & 0&.071&  0&.007 & 0&.102 & 0&.013 & 0&.0842\\
 3YY  & 0&.031&  0&.004 & 0&.044 & 0&.006 & 0&.0263\\
 3XZ  & 0&.045&  0&.001 & 0&.027 & 0&.020 & 0&.0424\\
 3YZ  &-0&.010&  0&.005 &-0&.009 & 0&.005 &-0&.0109\\
\end{tabular}
\end{table}

%
%
\begin{table}
\caption{The parameters of the shell model used for CaF$_2$. 
	 The notation follows 
	 Ref.~\protect\onlinecite{Elcombe70} and the ionic charge
	 Z(Ca) was fixed at $2e$ (shell model III in 
	 Ref.~\protect\onlinecite{Elcombe70}).
	}
\label{table_2}
\begin{tabular}{llr@{}l@{${}\pm{}$}r@{}lr@{}l@{${}\pm{}$}r@{}l}
  & Units & \multicolumn{4}{c}{RMC of PDF} &
               \multicolumn{4}{c}{Ref.~\protect\onlinecite{Elcombe70}}\\
\tableline
 $A_1$        & $e^2/v$ & 16&.00  &  0&.25  & 15&.12  & 0&.17 \\
 $B_1$        &         & -1&.84  &  0&.09  & -1&.70  & 0&.08 \\
 $A_2$        &         &  1&.30  &  0&.11  &  1&.30  & 0&.11 \\
 $B_2$        &         &  0&.09  &  0&.01  &  0&.09  & 0&.03 \\
 $A_3$        &         &  0&.22  &  0&.02  &  0&.23  & 0&.17 \\
 $B_3$        &         & -0&.28  &  0&.03  & -0&.28  & 0&.06 \\
 $A_4$        &         & -0&.18  &  0&.02  & -0&.18  & 0&.04 \\
 $B_4$        &         &  0&.062 &  0&.006 &  0&.061 & 0&.014\\
 $\alpha$(Ca) & $e$     &  1&.66  &  0&.16  &  1&.63  & 0&.11 \\
 $d$(Ca)      & $\AA^3$ & -0&.25  &  0&.02  & -0&.25  & 0&.03 \\
 $\alpha$(F)  & $e$     &  0&.45  &  0&.04  &  0&.45  & 0&.04 \\
 $d$(F)       & $\AA^3$ &  0&.065 &  0&.006 &  0&.063 & 0&.015\\
\end{tabular}
\end{table}


\end{multicols}

\end{document}